# An iterative wave-front sensing algorithm for high-contrast imaging systems[*]

**Jiangpei Dou,**[1,2] **Deqing Ren,** [1,2,3] **and Yongtian Zhu**[1,2]

[1] *National Astronomical Observatories / Nanjing Institute of Astronomical Optics & Technology, Chinese Academy of Sciences, Nanjing 210042, China*

[2] *Key Laboratory of Astronomical Optics & Technology, Nanjing Institute of Astronomical Optics & Technology, Chinese Academy of Sciences, Nanjing 210042, China*

[3] *Physics & Astronomy Department, California State University Northridge, 18111 Nordhoff Street, Northridge, California 91330–8268, USA*

**Abstract** Wave-front sensing from focal plane multiple images is a promising technique for high-contrast imaging systems. However, the wave-front error of an optics system can be properly reconstructed only when it is very small. This paper presents an iterative optimization algorithm for the measurement of large static wave-front errors directly from only one focal plane image. We firstly measure the intensity of the pupil image to get the pupil function of the system and acquire the aberrated image on the focal plane with a phase error that is to be measured. Then we induce a dynamic phase to the tested pupil function and calculate the associated intensity of the reconstructed image on the focal plane. The algorithm is to minimize the intensity difference between the reconstructed image and the tested aberrated image on the focal plane, where the induced phase is as the variable of the optimization algorithm. The simulation shows that the wave-front of an optics system can be theoretically reconstructed with a high precision, which indicates that such an iterative algorithm may be an effective way for the wave-front sensing for high-contrast imaging systems.

**1. Introduction**

Discovering life on another planet will potentially be one of the most important scientific advances of this century. The search for life requires the ability to detect photons directly from one planet and the use of spectroscopy to analyze its physical and atmospheric conditions. The direct imaging of an Earth-like low-mass planet orbiting its bright primary star is, however, extremely challenging. For NASA's Terrestrial Planet Finder Coronagraph, a contrast of $10^{-10}$ at an inner working angular (IWA) distance better than $4\lambda/D$ is required in the visible wavelength (Brown & Burrows 1990). Recently many high-contrast coronagraphs have been proposed for the direct imaging of an Earth-like exoplanet which can theoretically reach $10^{-10}$ contrast at a few $\lambda/D$ from a bright star (Ren &Serabyn 2005; Guyon et al. 2006; Ren &Zhu 2007). However, most of existing coronagraphs can only reach a contrast in the order of $10^{-5} \sim 10^{-7}$, even in the laboratory (Kasdin et al. 2004; Dou et al. 2010; Ren et al. 2010). One of the main contrast limitation factors comes from the wave-front error induced by imperfections in both the telescope optics and the coronagraph, which forms static speckle noises surrounding the star image (Ren & Wang 2006). The local speckles are much brighter than the planet image because of the large bright difference between the planet and its parent star, thus making the direct imaging of an Earth-like exoplanet impossible. For the space-based observation, the wave-front error changes very slowly and high S/N can be achieved by increasing the exposure time. For ground-based observation, the static wave-front error is one of the major error sources, although dynamic wave-front error that is induced by the atmospheric turbulence and is changing from time to time dominates the imaging performance. For an extreme adaptive optics system dedicated to the direct imaging of an exoplanet, both errors need to be corrected efficiently. In Serabyn's recent paper, a phase retrieval wave-front sensing algorithm has been proposed to correct the static speckle and three planets around the star of HR 8799 have been imaged (Serabyn et al., 2010).

[*] Supported by the National Natural Science Foundation of China

To remove the wave-front error induced speckles, one promising approach is called the speckle nulling technique, which can be realized by using a deformable mirror (DM) to induce phase on the pupil plane of the coronagraphic system. With a specific phase provided by the DM, it can theoretically create a local high-contrast zone on the focal plane that can be served as a discovery area for the exoplanet direct imaging (Malbet 1995). Based on the speckle removing technique, an extra-contrast improvement of $10^{-2} \sim 10^{-3}$ is expected to gain after eliminating the speckle noise surrounding the star image.

The most important procedure for the high-contrast zone correction is to precisely measure the wave-front error of the system. In recent papers, different algorithms were presented to reconstruct the wave-front directly from multi-images taken on the focal plane of an optics system. Firstly, these algorithms can work only when the original wave-front error of an optics system is not too large, due to the approximation that was introduced on either the phase error to be measured or the phase provided by the DM (Borde & Traub 2006; Give'on et al. 2007; Dou et al. 2009). Secondly, the DM deformation must be chosen carefully to create multi-images uncorrelated to each other, so that the denominator "D" in above algorithms will not be zero. Otherwise, there will be no solution for the above algorithm, in which case the wave-front can never be properly reconstructed. Thirdly, in the above algorithm the phase provided by the DM is calculated through the so called influential function model, which may further induce unnecessary errors in the actual process of the wave-front sensing.

To overcome these problems, this paper presents an iterative optimization algorithm for wave-front measurement directly from one single focal plane image. Firstly, the intensity of the image taken on the pupil plane of the optics system is measured and its electric field magnitude can directly be calculated. Then we introduce a dynamic phase to the pupil and perform the Fourier transformation to calculate the associated intensity of the reconstructed focal plane image. Such a reconstructed image will be compared with the actual focal plane image that includes the actual wave-front error of the optics system. The algorithm is to minimize the intensity difference between the reconstructed image and the actual focal plane image of the optics system, where the induced wave-front error is as the variable. In each step of the iterative optimization procedure, the induced phase, as the only variable of the optimization algorithm, will change to make the intensity of the two images approximating to each other. In the whole procedure, there is no approximation on the phase, which may guarantee a high precision for the measurement of relatively larger wave-front errors. For demonstration purpose, a numerical simulation is performed based on the graphical user interface (GUI) of the Optimization Tool in Matlab.

Simulation results show that the wave-front of an optics system can be theoretically reconstructed with a high precision, which shows that such an iterative algorithm may be an effective way for the wave-front sensing for a high-contrast imaging system.

The outline of the paper is as following. In Section 2, the principle of the wave-front sensing algorithm is proposed. In Section 3, we present the numerical simulation of the iterative optimization algorithm. The summary and conclusions are given in Sections 4.

## 2. PRINCIPLE OF THE WAVE-FRONT SENSING ALGORITHM

Recent laboratory tests have demonstrated that the actual coronagraph can provide a contrast in the order of $10^{-5} \sim 10^{-7}$. Further improvement is limited by the speckle noise that is induced from the wave-front error of the coronagraphic system. In this section, we consider a general optics system with a circular entrance pupil. For simplicity, the system will be operated at a monochromatic wavelength.

For an optics system with wave-front errors, the electric field of the electromagnetic wave at the pupil plane of the system can be expressed:

$$E_{pupil}(u,v) = A(u,v)e^{i\phi(u,v)}, \qquad (1)$$

where $A(u,v)$ represents the pupil function of the optics system; $\Phi(u,v)$ represents the original wave-front or phase error of the optics system that will induce spot-like speckles surrounding the bright star image on the focal plane.

We firstly put a CCD camera on the pupil plane of the optics system and measure the intensity of the pupil image. Then the magnitude of the electric field on the pupil or the so called pupil function can be achieved directly from such intensity and is given as:

$$A_t(u,v) = \sqrt{I_{pupil}}, \qquad (2)$$

where $I_{pupil}$ represents the tested intensity of the pupil image.

Then we induce a dynamic phase to the tested pupil function and the reconstructed electric field on the pupil can be expressed:

$$E_r(u,v) = A_t(u,v)e^{i\Psi(u,v)}, \qquad (3)$$

where $\psi$ is the induced dynamic wave front.

Since the starlight is much brighter than that of the planet, the planet image that is much less in intensity than the star image is negligible during the wave-front sensing process without causing any significant error. The electric field of the starlight on the focal plane is the Fourier transform of the aberrated electric field on the pupil plane of the system. Then the reconstructed electric field on the focal plane can be expressed as:

$$E_{focal\_r}(x,y) = \vec{F}[E_r(u,v)], \qquad (4)$$

where $\vec{F}$ represents the Fourier transform of the associated function.

The point spread function (PSF) of the starlight on the focal plane is square of the complex modulus of the aberrated electric field. The intensity of the reconstructed PSF image is given as:

$$I_r(x,y) = \left|E_{focal\_r}(x,y)\right|^2. \qquad (5)$$

Combining Equation (3), (4) and (5), the intensity of the reconstructed focal plane image becomes:

$$I_r(x,y) = \left|\vec{F}[A_t(u,v)e^{-i\Psi(u,v)}]\right|^2. \tag{6}$$

The principle of this algorithm is totally different from previously proposed ones (Borde & Traub 2006; Give'on et al. 2007; Dou et al. 2009). In the iterative algorithm, only one focal plane image rather than 3 images is used to reconstruct the original wave-front. No approximation is introduced on the original phase error, making it suitable for the measurement of much larger wave-front errors with a high precision. Here we put the CCD camera to measure the intensity of the starlight image on the focal plane of the actual optics system and it can be expressed as:

$$I_t(x,y) = \left|\vec{F}[A(u,v)e^{-i\phi(u,v)}]\right|^2, \tag{7}$$

where $I_t$ is the tested intensity of the starlight image on the focal plane of the actual system; $\Phi$ is the actual phase error of the optics system that is to be measured.

To measure the original wave-front error of the optics system, the optimization algorithm is to minimize the intensity difference between the reconstructed image and the actual focal plane image taken on the CCD. Here we subtract the intensity of each pixel on the two images and acquire the sum of the absolute value of the residual intensity. Once the sum of residual intensity becomes a minimum, the intensity of the two images will be approximated to each other. As a result, the value of the induced phase will be nearby the actual phase error that is to be measured. The optimization algorithm is to minimize the following equation:

$$\min\left\{\sum_{x=1}^{M}\sum_{y=1}^{N}|I_r(x,y) - I_t(x,y)|\right\}, \quad subject\ to \quad -\lambda/2 \leq \Psi \leq \lambda/2, \tag{8}$$

assuming the CCD with MxN pixels.

Substituting Equation (6) and (7) in Equation (8), the problem has become to minimize:

$$\min\left\{\sum_{x=1}^{M}\sum_{y=1}^{N}\left|\left|\vec{F}[A_t(u,v)e^{-i\Psi(u,v)}]\right|^2 - \left|\vec{F}[A(u,v)e^{-i\phi(u,v)}]\right|^2\right|\right\}. \tag{9}$$

Here in this paper, we use the Zernike polynomial to represent both the original phase error and the induced phase:

$$\Phi = \sum_{n=1}^{N} a_n Z_n; \quad \Psi = \sum_{n=1}^{N} a'_n Z_n, \tag{10}$$

where $a_n$ and $a'_n$ are the Zernike coefficients; and N is the order of the Zernike polynomial of $Z_n$; because each order of the Zernike polynomial is uncorrelated with each other, it can guarantee the existence of a solution for Equation (9).

In the procedure of optimization, we provide an initial phase (for instance ψ=0), as the start of the iterative algorithm (the first step). In each iterative step, the induced phase ψ will change towards the direction that makes the residual intensity become smaller. With a reasonable initial phase, the wave-front can be reconstructed very quickly in several steps. Once the wave-front error of the optics system can be precisely measured, it may be corrected by inducing an appropriated DM and the speckle noise surrounding the bright starlight image will be expected to be effectively eliminated. A numerical simulation based on the iterative algorithm will be presented in detail in the next section.

### 3. NUMERICAL SIMULATION

The feasibility of the wave-front sensing algorithm discussed above can be verified by the following numerical simulation. For demonstration purpose, we create a distorted optics system by introducing a random phase error to an ideal optics system with a circular pupil. Here the root mean square (RMS) of the phase error is ~ 0.2114 rad, which is to be measured. And the aberrated image on the focal plane will be as the reference image that is to be compared with the reconstructed image. In this paper we use a 36-order Zernike polynomial to represent the phase and the 36 Zernike coefficients are created randomly, for a general purpose. Since each order of the Zernike polynomial is uncorrelated with each other, it can guarantee the existence of a solution for the iterative optimization algorithm. The amplitude pattern of the pupil function (image taken on the pupil), the theoretical and aberrated PSF (the starlight image on the focal plane) of the system are shown in Figure 1.

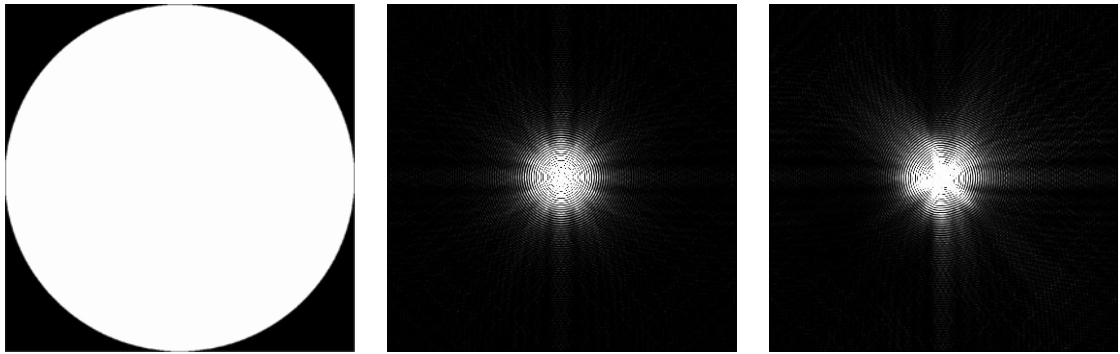

**Fig. 1** Left: The amplitude pattern of the pupil function; Middle: The theoretical PSF of the starlight image; Right: The aberrated image with a phase error that is to be measured.

In the procedure of the iterative optimization, a dynamic phase, as the only variable of the optimization problem, will be induced to the theoretical pupil function and the intensity of associated PSF can be calculated directly by using the Fourier transformation (here in this simulation, we use a 2-D fast Fourier transformation). In each iterative step, such a calculated intensity that changes with the induced phase will be subtracted from the intensity of the aberrated image with original phase error (the referenced image). In the next step, the phase will change towards the direction that makes residual intensity become smaller. Once the residual intensity becomes a minimum, the iterative procedure will stop and the phase in the last step will be the optimum phase that best approximates the original one.

Such a problem has become a 2-D constrained nonlinear minimization problem. Here the algorithm is based on the GUI of the Optimization Tool in Matlab. The object function of such an optimization algorithm is to find a minimum of residual intensity between the reference image and the reconstructed image. Supposing the wave-front error of the system is within a wavelength and the constraint function of the algorithm will be represented as $-\lambda/2 \leq \psi(u,v) \leq \lambda/2$, for the phase that is induced to the pupil plane. The procedure of the optimization algorithm will take several steps and converge very quickly provided with a reasonable start point. Here we initialize the start point to be $\psi(u,v)=0$ for a general purpose. Figure 2 shows the original phase map to be measured and the reconstructed wave-front map by using the algorithm, respectively. It clearly indicates that the reconstructed wave-front is very consistent with the original one based on the algorithm.

A trade-off is needed between the accuracy of the wave-front sensing and the velocity of the procedure. For example, during the iterative procedure the convergence will be very fast for the first 50 steps with a remaining phase error of RMS in the order of $10^{-2}$ rad. However, the convergence velocity will greatly decrease when the remaining phase error is very small. That means we have to wait for a long time if an extremely high precision for the wave-front sensing is needed. Here we stop the optimization procedure after 86 steps with an acceptable accuracy. The reconstructed wave-front is of a RMS ~ 0.211485 rad, with a remaining wave-front error in the order of $10^{-3}$ rad (RMS). Comparing with original wave-front, the relative error is 0.004%. Figure 3 shows the remaining wave-front error and the residual intensity between the reconstructed image and the aberrated starlight image of the actual optics system.

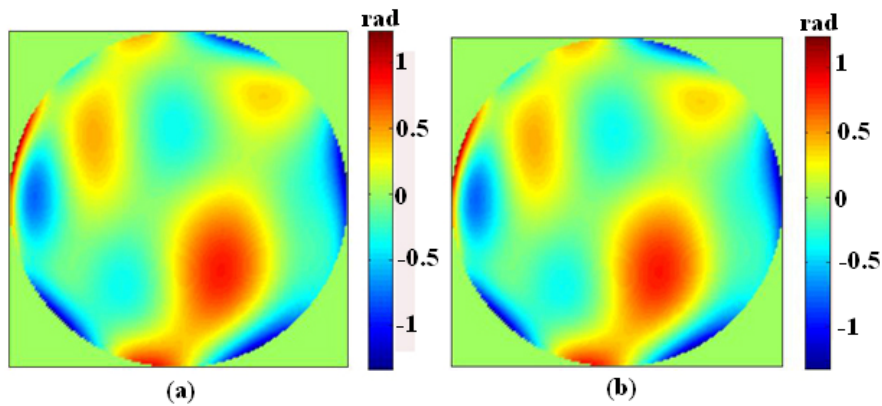

**Fig. 2** The original wave-front map that is to be measured (left) and the reconstructed wave-front map based on the iterative optimization algorithm (right).

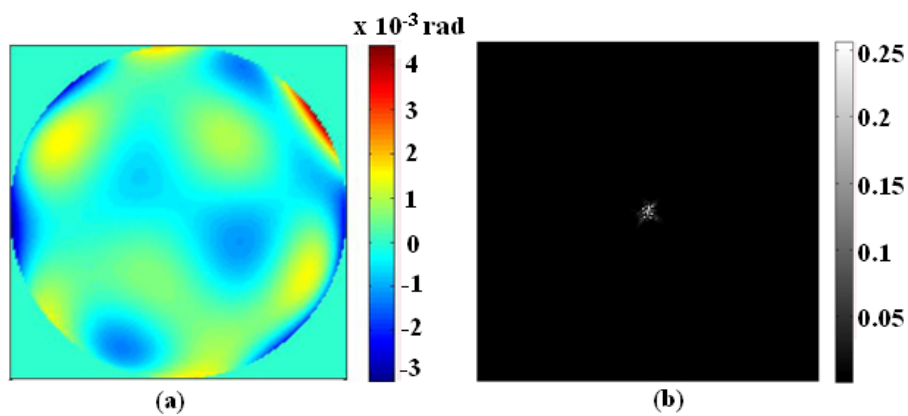

**Fig. 3** The remaining wave-front error (left) and the residual intensity between two starlight images (right)

Since the original wave-front error is created randomly, the algorithm should be suitable for a general situation. To testify that, we also use other wave-front maps to replace the one that has been used above, and achieve the same result, which gives us confidence that the performance of the iterative optimization algorithm is reliable and promising for the wave-front measurement for high-contrast imaging systems.

**4 SUMMARY AND CONCLUSIONS**

Based on the iterative optimization algorithm, the wave-front error of an optics system can be precisely measured directly from one focal plane image, which has been demonstrated in our numerical simulation. The iterative optimization algorithm we have proposed here is totally different from the multi-image focal plane wave-front sensing algorithms. Since no approximation is performed on the original wave-front error or on the DM induced phase during the whole procedure, such an algorithm can be used for the measurement of large wave-front errors, which is impossible for the previously proposed 3-image focal plane wave-front sensing algorithm. Although in this paper we have only considered for a monochromatic wavelength, for a general coronagraphic system, wave-front at other wavelength can be simply scaled according to the actual wavelength. At present a laboratory test system has been set up for such a wave-front sensing. We will discuss the later result in the future publication.

**Acknowledgements**

We acknowledge the anonymous referee for his/her thoughtful comments and insightful suggestions that improved this paper greatly. This work was funded by the National Natural Science Foundation of China (NSFC) (Grant No. 10873024) and was partially supported by the National Astronomical Observatories' Special Fund for Astronomy. Part of the work described in this paper was carried out at California State University Northridge, with support from the National Science Foundation under grant ATM-0841440.